\documentclass[aps,prl,twocolumn,superscriptaddress,showpacs]{revtex4}
\usepackage{graphicx}

\newcommand{\msr}{$\mu$SR}
\begin{document}



\title{Muonium as a shallow center in GaN }


\author{K. Shimomura}
\affiliation{Institute of Materials Structure Science, High Energy Accelerator Research Organization (KEK), Tsukuba, Ibaraki 305-0801, Japan}
\author{R. Kadono}
\altaffiliation[Also at ]{ School of Mathematical and Physical Science,
The Graduate University for Advanced Studies}
\affiliation{Institute of Materials Structure Science, 
High Energy Accelerator Research Organization (KEK), Tsukuba, Ibaraki 305-0801, Japan}
\author{K. Ohishi}
\affiliation{Institute of Materials Structure Science, 
High Energy Accelerator Research Organization (KEK), Tsukuba, Ibaraki 305-0801, Japan}
\author{M. Mizuta}
\altaffiliation[Present address: ]{Photonic and Wireless Devices Research Laboratories, 
NEC Corpration, Ohtsu, Shiga 520-0833, Japan}
\affiliation{System Devices and Fundamental Research Laboratories, NEC Corporation, 
Tsukuba, Ibaraki 305-8501, Japan}
\author{M. Saito}
\altaffiliation[Present address: ]{Computational Materials Science Center, 
National Institute for Materials Science, Tsukuba, Ibaraki 305-0047, Japan}
\affiliation{System Devices and Fundamental Research Laboratories, 
NEC Corporation, Tsukuba, Ibaraki 305-8501, Japan}
\author{K. H. Chow}
\affiliation{Department of Physics, University of Alberta, 
Edmonton, Alberta, Canada T6G 2J1}
\author{B. Hitti}
\affiliation{TRIUMF, 4004 Wesbrook Mall, Vancouver, Canada V6T 2A3}
\author{R. L. Lichti}
\affiliation{Department of Physics, Texas Tech University, 
Lubbock, Texas 79409-1051, USA}


\date{\today}

\begin{abstract}
A paramagnetic muonium (Mu) state with an extremely
small hyperfine parameter was observed for the first time 
in single-crystalline GaN below 25 K.
It has a highly anisotropic hyperfine structure with 
axial symmetry along the $\langle$0001$\rangle$ direction, 
suggesting that it is located either at a nitrogen-antibonding or
a bond-centered site oriented parallel to the c-axis.
Its small ionization energy ($\leq 14$ meV) and small 
hyperfine parameter ($\sim10^{-4}$ times the vacuum value) indicate 
that muonium in one of its possible sites produces a shallow state, 
raising the possibility that the analogous hydrogen center 
could be a source of $n$-type conductivity in as-grown GaN.
\end{abstract}


\maketitle




Since the discovery of methods to produce sufficient $p$-type conductivity by 
Mg-doping \cite{amano:89,nakamura:92}, gallium nitride  and 
related compound semiconductors are being aggressively developed
for electronic and optelectronic devices  such as blue/green 
lasers and light-emitting diodes.  Unique features such as a wide and 
direct band gap and high breakdown field make the nitrides 
ideal for such applications. However, as-grown undoped GaN epitaxial 
thin films, as well as  bulk single crystals, 
commonly exhibit $n$-type conductivity with concentrations 
ranging  from $10^{16}$ to $10^{19}$ cm$^{-3}$.
 Extensive experimental and theoretical studies \cite{manasreh:00}
have been undertaken to understand
the origin of this $n$-type conductivity. 

For many years, nitrogen vacancies, which are commonly observed 
in as-grown GaN, were thought to be a major source of the $n$-type 
conductivity \cite{maruska:69,ilegems:73}.  
Recent theoretical work \cite{neu:94} challenges this view by showing 
that the nitrogen vacancies have a high formation energy in $n$-type GaN; 
hence their concentration is predicted to be too low to affect 
the electrical conductivity. 
Moreover, Hall effect measurements on a GaN sample 
irradiated by high-energy electrons showed that the donor 
level of the nitrogen vacancies (64(10) meV) 
was much deeper than that of the residual donors (18 meV) \cite{look:97}. 
Contamination by  oxygen or silicon is difficult to avoid 
during crystal growth and both of these impurities have been 
proposed as the origin of unintentional $n$-type behaviour.
Silicon is often used as an intentional donor dopant and can provide 
free electron concentrations of up to $10^{20}$\,cm$^{-3}$. 
Very recent magneto-optical studies of 
hydride-vapor-phase epitaxial GaN 
show that a candidate other than O or Si impurities can 
also act  as a shallow donor \cite{moore:02}.

By contrast, magnesium is currently the only acceptor dopant  
reliably  used  to obtain  $p$-type GaN. 
Even in this case,  high hole concentrations  were difficult 
to achieve  until it was found that hydrogen reacts with and passivates 
the Mg acceptors \cite{vech:92},
similar to its effect  on acceptors in Si. 
A post-growth anneal at  high temperature 
is required to remove H from the vicinity of the Mg dopants 
and thereby activate $p$-type electrical properties.

Motivated by such a crucial role of hydrogen in GaN, extensive 
\msr\ (muon spin resonance) studies \cite{chow:00} have been 
performed to clarify the physical and electronic structure 
of isolated H centers {\it via} their muonium analog.  
The results established the existence of two  
charged  states (Mu$^+$ and Mu$^-$) \cite{lichti:00, lichti:01}, 
and level-crossing resonance spectra  
revealed that Mu$^+$ and Mu$^-$ reside at  sites anti-bonding 
to  N  and  Ga, respectively \cite{lichti:03}.

In addition, neutral  paramagnetic muonium states, i.e. Mu$^0$, 
are readily observed in a wide variety of semiconductors. 
While the dynamical aspects (e.g., diffusion) 
may be considerably different between Mu and H due to the 
light mass of Mu ($m_{\mu} \simeq\frac{1}{9} m_p$), 
the local electronic structure of Mu is virtually equivalent 
to that of H after a small correction due to the difference 
in the reduced mass ($\sim4$\%). 
Recently, a novel Mu state having an extremely small hyperfine 
parameter ($10^{-4}\times A_\mu$) was reported in II-VI 
compound semiconductors including CdS \cite{gil:99} 
and ZnO \cite{cox:01, gil:01, shi:02}, implying that  
Mu (and hence H) can act as a donor in these materials.
No Mu$^0$ hyperfine spin-precession spectra 
have previously been reported for GaN,
 although hyperfine decoupling measurements suggest a
short-lived atomic-like neutral \cite{lichti:03}.

In this Letter we report on the first observation of a 
paramagnetic muonium spectrum in GaN.  The observed Mu$^0$ 
state has an extremely small 
and highly anisotropic hyperfine parameter. 
The location within the band gap for the [0/+] 
energy level associated with this Mu state is estimated 
from the measured  activation  energy 
for thermal ionization. These results imply that 
an isolated hydrogen impurity would behave as a shallow donor 
if it were  located at the same crystallographic site.

We performed \msr\  measurements on a GaN single-crystal 
with the hexagonal (2H) wurtzite structure.  GaN has tetrahedral 
coordination with bonds parallel to the symmetry axis slightly 
elongated  and  off-axis bond directions  ordered to give
hexagonal symmetry around [0001].  
The \msr\ experiment was conducted at the TRIUMF 
M15 beamline with the Belle spectrometer. Muons from a surface beam 
(100\% spin-polarized  with an energy of 4 MeV) 
with their polarization transverse to the applied magnetic field 
were implanted into GaN 
(3\,$\times$\,3\,$\times$\,0.5\,mm, [0001] orientation, 
$n$ type with a  concentration of $10^{16}$\,cm$^{-3} $).
The subsequent time evolution of the muon spin precession yields 
information on the muonium hyperfine parameters and  stability 
of the center.  High  fields (up to 5 T) were used to quench
fast spin relaxation due to interactions with host nuclear spins.
The GaN crystal was mounted in a He gas-flow cryostat such that the [0001] 
and [11$\bar{2}$0] crystallographic axes could be oriented at
specific angles  to the applied  field in order to 
examine the  symmetry of the hyperfine interaction.
Dependences on temperature and magnetic field strength 
were obtained with the field parallel to the [0001] axis.

\begin{figure}[t]
\includegraphics[width=0.9\linewidth]{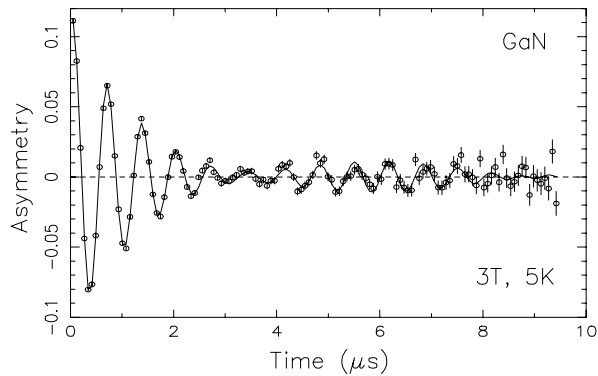}
\caption{The \msr\ time spectrum in GaN at 5.0 K under
an external field of 3.0 T applied parallel to the [0001] axis, displayed in a
rotating-reference-frame frequency of 405 MHz. A beat pattern
due to multiple frequencies is clearly seen.}
\label{fig1}
\end{figure}

Above 25 K, only a single (diamagnetic) precession signal is observed
at the muon Larmor frequency 
(gyromagnetic ratio $\gamma_\mu=2\pi\times135.53$ MHz/T).
Relaxation of this signal is well described by Gaussian damping
with a nearly temperature-independent rate constant of 
$\simeq0.2$ $\mu$s$^{-1}$.   This damping rate is satisfactorily
explained by the dipole-dipole interaction of muons with 
$^{69,71}$Ga and $^{14}$N nuclei. The muon spin rotation 
signal changes drastically below 25 K. 
A typical \msr\ spectrum is shown in Fig.~1. 
Fig.~2 shows the angular and temperature dependence of the 
frequency spectra obtained by Fourier transform, 
in which one pair of satellite lines is clearly seen 
with their positions situated symmetrically around the central line, 
which corresponds to the precession of diamagnetic muons.
The splitting of these satellites remained unchanged when 
the applied field was increased from 1.5 T up to 5 T (see Fig.~2a and 2c), 
a result that is important in identifying the spectra as due to 
the hyperfine interaction of a Mu$^0$ center.  
The splitting decreases when the [0001] axis 
is tilted with respect to $\vec{B}$ as in Fig.~2d.
Moreover, an  equivalent frequency spectrum was observed
when the [11$\bar{2}$0] axis was rotated by 90$^\circ$ around [0001], 
which was oriented at 35$^\circ$ to the applied magnetic field. 
These  observations demonstrate the presence of a paramagnetic 
muonium state in GaN.  The resulting hyperfine interaction is 
extremely small, about $10^{-4}$ times the vacuum value for a Mu atom,  
and is axially symmetric with respect to [0001].

\begin{figure}
\includegraphics[width=0.9\linewidth]{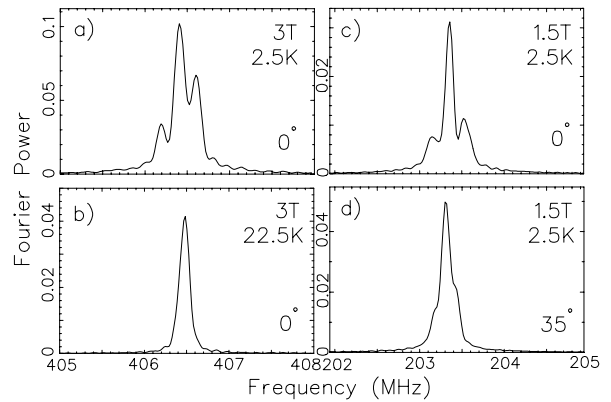}
\caption{Frequency spectra obtained for GaN at (a) 2.5 K and (b) 22.5 K 
with $B$=3.0 T parallel to [0001] axis, and with $B$=1.5T at 2.5 K,  
where [0001] is parallel to $B$ (c) or tilted by 35$^\circ$ 
from $B$ (d), where from 0.01 $\mu$s to 9.60 $\mu$s is adopted as a time range of FFT.
The number of y-axis in (a) and the other is not directly comparable, because of the difference in the data statistics. }
\label{fig2}
\end{figure}


With the [0001] axial symmetry established qualitatively, a more formal
analysis can be undertaken to extract the hyperfine constants. 
The amplitude differences for the two hyperfine lines 
allow the specific muonium transition to be assigned
to each satellite and establishes the sign of the hyperfine constants.
Spin precession signals from the paramagnetic Mu state have two
components in the high-field limit ($B\gg 2\pi A/\gamma_e$, where 
$\gamma_e=2\pi\times28.024$ GHz/T is the electron gyromagnetic ratio):
\begin{eqnarray}
\nu_{12}(\theta)&\simeq&\nu_0-\frac{1}{2}\Delta\nu(\theta),\\
\nu_{34}(\theta)&\simeq&\nu_0+\frac{1}{2}\Delta\nu(\theta),\\
\Delta\nu(\theta) &=& A(\theta) = A_{\parallel }\cos^2\theta +
A_{\perp }\sin^2\theta,
\end{eqnarray}
where $2\pi\nu_0=\gamma_\mu B$,  $\theta$ is the angle between 
$\vec{B}$ and the [0001] symmetry axis, and $ A_{\parallel}$ and 
$ A_{\perp}$ are the hyperfine parameters parallel and perpendicular 
to [0001], respectively. The frequencies $\nu_{12}$ and $\nu_{34}$ 
correspond to the transitions of spin states between
${\mid}s_e,s_{\mu}{\rangle}={\mid}+\frac{1}{2},+\frac{1}{2}{\rangle}$ and 
${\mid}+\frac{1}{2},-\frac{1}{2}{\rangle}$,  
and  between ${\mid}-\frac{1}{2},+\frac{1}{2}{\rangle}$ and 
${\mid}-\frac{1}{2},-\frac{1}{2}{\rangle}$, respectively. 

\begin{figure}
\includegraphics[height=0.6\linewidth]{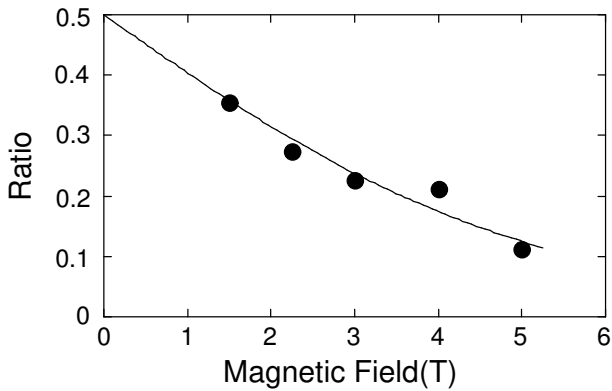}
\caption{Magnetic field dependence of the ratio of the lower-frequency satellite amplitudes to the sum of both satellites at 2.5 K (symbols). 
Solid curve is a fitting result using the Maxwell-Boltzmann distribution 
with the Mu temperature as a free parameter.}
\label{fig3}
\end{figure}

Because the population of the $s_e=-\frac{1}{2}$ state is always larger in an applied field,
the observations in Fig.~2 imply that the low-frequency line
corresponds to $s_e=+\frac{1}{2}$, thus to $\nu_{12}$, and that $A(\theta)$ 
is positive for the displayed orientations.
Results from direct fits to the time-domain spectra (see Fig.~1), where 0.01 $\mu$s to 9.60 $\mu$s in time range, provide a better amplitude measurement for 
comparison with theoretical expectations.  Analysis assuming  
three frequency components  ($\nu_0$, $\nu_{12}$, and  $\nu_{34}$, we assumed the depolarization rates of the two Mu lines are equal )
is quite satisfactory, with a reasonably small reduced 
$\chi^2$ ($\simeq1.55$) for the data in Fig.~1.  Fig.~3 shows the  
field dependence for the ratio of the lower-frequency satellite amplitudes to the sum of both satellites at 2.5\,K and $\theta = 0$, together with  a fitting result of 
electron spin occupation probabilities for 
an isolated muonium center unsing the Maxwell-Boltzmann distribution, 
where 3.4\,K was obtained as the Mu temperature. 
Correspondence between the sample and obtained Mu temperature 
is also satisfactory, confirming assignment of a positive $A_{\parallel}$ 
hyperfine constant.

From the spectrum in Fig.~2a, with $\vec{B}$ applied along [0001], 
the hyperfine constant is  deduced to be
\begin{eqnarray}
A (0^\circ) &=& A_{\parallel } = +337(10)\:{\rm kHz}.
\end{eqnarray}
Combining this result with the data for a tilted sample orientation 
with respect to $\vec{B}$ 
(Fig.~2, where $\theta = 35.0^{\circ}$, and $\Delta\nu = 146(3)$kHz) 
the remaining hyperfine constant is found to be
\begin{eqnarray}
A_{\perp} &=& -243(30)\:{\rm kHz}.
\end{eqnarray}
As an experimental check, these parameters 
reproduce  the observed splitting for an additional
orientation with $\theta = 15.0^{\circ}$.  Therefore, we conclude that 
 the hyperfine parameters parallel and  perpendicular 
to the  [0001] symmetry axis have opposite signs.
The static dielectric constants in GaN are 
 10.4 parallel  and 9.5 perpendicular to [0001] \cite{manasreh:00}.
Since the degree of  anisotropy in the hyperfine 
tensor for the observed paramagnetic muonium state is much larger than 
that implied by the  dielectric constant, we conclude that 
the observed anisotropy is due to local site symmetry.
Of the likely locations  for Mu in the GaN wurtzite structure 
\cite{neu:95,wright:99,myers:00}, these results strongly suggest that the 
paramagnetic  Mu center resides at either the 
${\rm AB}_{N\parallel}$ or the ${\rm BC}_{\parallel}$ site.

\begin{figure}
\includegraphics[height=0.7\linewidth]{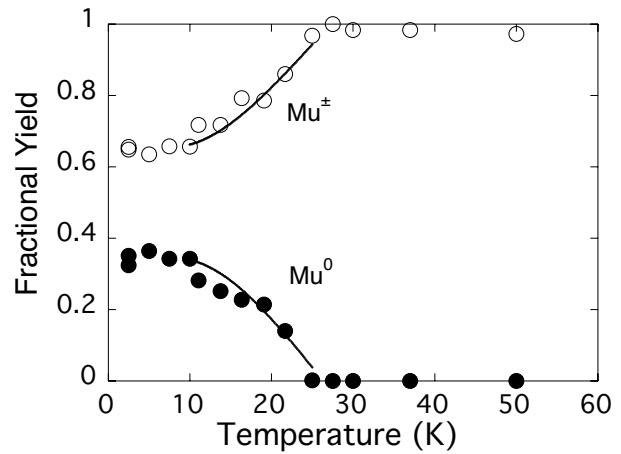}
\caption{The fractional yield of Mu (closed circles),
and a diamagnetic state (open circles) versus temperature in GaN. Solid curves are
fitting results (see text).}
\label{fig4}
\end{figure}

The temperature dependence of the amplitudes of the Mu$^0$  
and diamagnetic signals are plotted in Fig.~4. The total amplitude 
is  almost independent of temperature, 
suggesting that Mu  is ionized to a diamagnetic 
state above 25 K.   It seems unlikely that the 
energy levels are just above the valence band  
since, if that were correct, Mu should remain neutral
 due to the bulk $n$-type conductivity 
of the present specimen which puts the Fermi level much 
higher than mid-gap. These  results indicate that 
the observed  paramagnetic Mu center acts as a shallow donor. 
Thus, since Mu centers simulate the electronic structure of 
H in GaN, our result provides convincing evidence that  
hydrogen centers at this particular site are shallow donors, 
and thus would contribute to $n$-type conductivity in GaN.

The activation energy $E_a$ for Mu ionization was found 
to be 4.6(6) meV by fitting the data in Fig.~4
over a region from 10 K to 25 K with a function 
$\alpha+\beta\exp(-E_a/k_BT)$. Interpretation of
this energy depends on the precise process forming the neutral 
charge-state of the Mu donor and the extent to which 
equilibrium arguments apply to muonium state occupations.
At one end of the range it represents  a direct
transition from the defect level to the bottom of the
conduction band; while at the other extreme, in equilibrium
it is for a transition from the defect level to the Fermi energy.
Assuming minimum compensation, a reasonable estimate of 
$E_F$ at non-zero temperatures below the ionization of  
residual donors is midway between the 18 meV  level
for the dominant donors and the bottom of the 
conduction band.  Thus our $\sim$5 meV 
activation energy translates to a Mu[0/+] level somewhere 
between 5 and 14 meV below the conduction band edge. 
Considering the difficulty in interpreting the Mu 
ionization energy and ambiguity in determining the level
of the shallower low-concentration donor in
Ref.~\cite{moore:02}, the hydrogen equivalent of the
observed Mu$^0$ center might be a candidate for this
additional donor.

It is interesting  that for Mu/H in GaN,  no such shallow  
states are  predicted in the latest theoretical 
studies \cite{neu:95, walle:03}.   Our observation may
not be incompatible with those predictions if the muon
location that yields the shallow center were metastable 
rather than the lowest energy site.  Of the two
 sites that locally satisfy the hyperfine symmetry, 
BC$_{\parallel}$  is one 
of the lower-energy positions for both neutral and 
positive change states. Data for hydrogen at high temperatures
\cite{myers:00} indicate that both the AB$_{N\perp}$ and 
BC$_{\parallel}$ sites are occupied during H$^+$ diffusion.
The other likely site, AB$_{N\parallel}$, is 
 at a significantly  higher energy, but  yields 
a (meta)stable location for Mu$^+$ based on  
level-crossing results \cite{lichti:03}.
The same `cage' region of the GaN structure that 
contains AB$_{N\parallel}$  also contains the  
AB$_{Ga\parallel}$ site found for
a metastable Mu$^-$.  A  transition from Mu$^+$ to Mu$^-$ 
at these sites, observed above 150 K, must 
involve a short-lived neutral that remains
inside this cage. 
Related to the question of site,  
the origin for the high degree of anisotropy in the 
Mu hyperfine coupling is yet to be understood. 
A theoretical study of the shallow-donor  N-H complex 
 in diamond \cite{miyazaki:02}  
may prove helpful on this issue. 

More than 60\% of 
implanted muons form  diamagnetic states  
 even at the lowest temperature. 
Our data indicate that whenever the shallow Mu$^0$ is
present, the diamagnetic signal shows a fairly rapid 
exponential relaxation, characteristic of fast dynamics
such as a transition out of that state.  
One  possibility is that either Mu$^+$ or Mu$^-$, 
both of which were identified in the previous
experiments \cite{lichti:03}, represents the initial
state formed upon implantation and is a precursor to
slightly delayed formation of the shallow neutral
Mu center.   

Further experiments, including {\sl muonium} level 
crossing resonance measurements, are clearly required to 
clarify the detailed electronic structure and
to more precisely identify the site of the observed 
shallow Mu$^0$ center.

In summary, we have demonstrated that a paramagnetic 
muonium center with extremely small hyperfine parameters 
is formed in GaN below 25 K. This Mu$^0$ center has a 
hyperfine interaction that is axially symmetric  around [0001]. 
The temperature dependence of the  yield for this state 
indicates that it acts as a shallow donor, strongly suggesting 
that hydrogen, if located at an equivalent crystallographic site, 
would contribute to the unintentional $n$-type 
conductivity in GaN.

We would like to thank the staff of TRIUMF for their technical support. 
We are also grateful to K. Nagamine and K. Nishiyama for their continuous 
encouragement and helpful discussions.   This work was supported by  
the Natural Sciences and Engineering Researech Council of Canada (KHC, BH),
the US National Science Foundation (RLL), 
and the Robert A. Welch Foundation (RLL).

\end{document}